\documentclass[10pt,letterpaper]{article}
\usepackage[top=0.85in,left=2.75in,footskip=0.75in,marginparwidth=2in]{geometry}

\usepackage[utf8]{inputenc}
\DeclareUnicodeCharacter{221A}{$\surd$}

\usepackage{nameref,hyperref}

\usepackage[right]{lineno}

\usepackage{microtype}

\raggedright
\usepackage{changepage}

\usepackage[aboveskip=1pt,labelfont=bf,labelsep=period,singlelinecheck=off]{caption}
\usepackage{subcaption}

\usepackage{booktabs}
\usepackage{threeparttable}  

\makeatletter
\renewcommand{\@biblabel}[1]{\quad#1.}
\makeatother

\usepackage{lastpage,fancyhdr,graphicx}
\graphicspath{{./}} 
\usepackage{epstopdf}
\fancyheadoffset[L]{2.25in}
\fancyfootoffset[L]{2.25in}

\usepackage{color}
\usepackage{xcolor}

\definecolor{Gray}{gray}{.25}
\definecolor{sns-blue}{HTML}{4c72b0}
\definecolor{sns-orange}{HTML}{dd8452}
\definecolor{sns-red}{HTML}{c44e52}
\definecolor{sns-cyan}{HTML}{64b5cd}

\usepackage{graphicx}

\usepackage{sidecap}

\usepackage{wrapfig}
\usepackage[pscoord]{eso-pic}
\usepackage[fulladjust]{marginnote}
\reversemarginpar

\usepackage{gensymb}

\usepackage[numbers,sort&compress]{natbib}

\usepackage{tikz}

\begin{document}
\vspace*{0.35in}

\begin{flushleft}
  {\Large
    \textbf\newline{Revisiting urban heat indices in Switzerland using low-cost measurement networks}
  }
  \newline
  \\
  Mart\'i Bosch\textsuperscript{1,2,*},
  Moritz Burger\textsuperscript{1,2},
  \\
  \bigskip
  \bf{1} Oeschger Centre for Climate Change Research, University of Bern, Bern, Switzerland \\
  \bf{2} Institute of Geography, University of Bern, Bern, Switzerland
  \\
  \bigskip
  * marti.bosch@unibe.ch

\end{flushleft}

\section*{Abstract}

Urban populations are increasingly exposed to extreme heat events such as heatwaves, which can be exacerbated in cities due to the urban heat island (UHI) effect.
With the aim of developing adaptation strategies, recent years have seen a growing interest in deploying high-resolution measurement networks using low-cost devices (LCDs), which enable the evaluation of intra-urban temperature distribution and its impacts at an unprecedented spatial resolution.
However, the reliability of LCD measurements has been called into question, especially regarding potential overheating due to inadequate radiation shielding.

In this study, we develop a statistical method to correct temperature biases based on short-wave radiation using a generalized additive model (GAM) and then apply it to LCD measurements in the urban climate networks of the cities of Bern, Lausanne, Neuchatel and Zurich (Switzerland).
To that end, we first calibrate the correction procedure to the LCD models used in each city using an intercomparison field study, in which the LCD models are collocated next to a professional automated weather station (AWS) operated by MeteoSwiss in the rural surroundings of Bern. 
Then, we evaluate how these corrections can influence two climate indices, namely the number of tropical nights and the number of heat warnings issued in each city according to MeteoSwiss heat warning system.
The findings suggest that the current AWS underestimate the heat warnings, whereas some LCD models likely overestimate them due to radiative errors.
Nevertheless, uncorrected LCD measurements still provide a more reliable estimate of urban temperatures than AWS located outside urban settings.
The insights can guide selection of LCD models for new monitoring networks and support the application of model-specific radiative bias corrections to existing LCDs, enabling more accurate assessments of heat and its impacts.


\section*{Introduction}

Extreme heat is one of the most direct manifestations of climate change and a growing public health concern, contributing to excess mortality, reduced labour productivity and increased risk of heat-related illness \cite{phelan2015urban,santamouris2015impact,sera2019urban,casanueva2020escalating,santamouris2020recent,masselot2023excess}.
Cities are particularly vulnerable due to the urban heat island (UHI) effect, whereby urban temperatures raise above those of surrounding rural areas \citep{oke1982energetic}.
In Switzerland, observations from urban and rural AWS across Basel, Bern, Geneva, Lausanne and Zurich show that summer nighttime temperatures in city centers exceed rural levels by more than 2$\degree$C on average (with maxima reaching 6-7$\degree$C) \citep{gehrig2018stadtische,burgstall2019representing}.
With urban populations continuing to grow, assessing heat exposure at the local level is increasingly critical to identify vulnerable areas and support targeted adaptation strategies \cite{deilami2018urban,rosenzweig2018climate,hamdi2020state,masson2020urban}.

Heat-health warning systems translate meteorological observations into actionable public health alerts and have been widely adopted across Europe to reduce the impacts of extreme heat on the population \citep{casanueva2019overview}.
In Switzerland, MeteoSwiss operates a heat warning system using daily mean temperature as main indicator \citep{burgstall2021von}, informed by epidemiological evidence linking heat exposure to adverse health outcomes \citep{ragettli2023explorative}.
The number of tropical nights --- during which the minimum temperature does not drop below 20$\degree$C --- is another widely used heat indicator which is linked to health risks particularly for vulnerable groups such as the elderly, the chronically ill, and children \cite{vicedo2016excess,kollner2017klimabedingte,ragettli2017exploring}.
Tropical nights are directly tied to the UHI effect and have seen a notable increase in Swiss cities in recent decades \citep{gehrig2018stadtische,burgstall2019representing}.
Both indicators, however, are strongly sensitive to the location of the meteorological stations used as input --- yet the AWS networks feeding these systems are predominantly located outside urban cores.

Quantifying intra-urban temperature variability at the resolution needed for such assessments requires observation networks with a spatial resolution that traditional AWS networks are not designed to provide \cite{grimmond2006progress,oke2006initial,grimmond2010climate}.
Professional weather stations operated by national meteorological services are designed to meet World Meteorological Organization (WMO) siting standards, which are difficult to satisfy in urban environments, and are therefore predominantly located in suburban or rural settings \cite{world2024guide}.
As a result, while existing AWS networks may capture the city-wide UHI signal when paired with rural reference stations, they are ill-suited to resolve the fine-scale spatial patterns of urban heat exposure that are relevant for adaptation planning and public health assessment.
This limitation has motivated a growing interest in deploying dense networks of low-cost devices (LCDs) in cities, which can achieve the spatial coverage needed to characterise intra-urban temperature variability at an unprecedented resolution \cite{muller2013sensors,hamdi2020state,masson2020urban,wong2025government}.

However, the reliability of LCD temperature measurements has been called into question, particularly regarding radiation-driven warm biases caused by inadequate shielding of the thermometer from incoming shortwave radiation \citep{bell2015good,chapman2016can,buchau2018modelling,cornes2020correcting}.
These biases exhibit a clear diurnal pattern, with overestimation of temperature during daytime hours and near-unbiased readings at night, and their magnitude depends strongly on the sensor model and radiation shield design \citep{jenkins2014comparison,ahmed2025comparison}.
Several correction approaches have been proposed, modelling the bias as a function of incoming solar radiation and, in some cases, other meteorological variables such as wind speed \citep{cornes2020correcting,beele2022quality,anet2024improving}.
In Switzerland, \citet{gubler2021evaluation} documented substantial daytime biases in the Bern LCD network and identified radiative bias correction as a key avenue for improving data quality. 
Accordingly, \citet{burger2021modelling,burger2022modeling} subsequently restricted their intra-urban temperature models in Bern to nighttime precisely because of these uncorrected biases, while \citet{zumwald2021mapping} noted that radiation-induced biases in citizen weather stations in Zurich could not be fully removed by their quality control procedure.

In this study, we address this gap by evaluating five LCD models against a professional MeteoSwiss AWS in a controlled intercomparison field study, developing a GAM-based radiative bias correction calibrated to each model.
We then assess the impact of raw and corrected LCD measurements on two heat indices --- the number of tropical nights and MeteoSwiss heat warning levels --- across the urban climate networks of Bern, Lausanne, Neuchatel and Zurich.
By comparing heat indices derived from LCD and AWS data, we show that professional AWS located outside urban settings systematically underestimate the heat exposure experienced within cities.
Additionally, we show that this gap cannot be solely attributed to radiative biases of the LCDs.
Our findings provide practical guidance for the selection and deployment of LCD models in urban climate networks and demonstrate that, even for sensor models prone to radiative errors, a simple statistical correction enables more representative assessments of urban heat than reliance on rural AWS alone.

\section*{Results}

\subsection*{Intercomparison field study}

\subsubsection*{Agreement metrics between LCD models and the reference AWS}
\label{sec:agreement-metrics}

The results of the intercomparison field study suggest notable differences in the agreements between each LCD model and the reference AWS 
(\nameref{sec:code-agreement-metrics}).
As shown in \autoref{tab:agreement-metrics}, the Barani and Koalasense models show overall negative biases as indicated by the mean and median bias errors (MBE and MdBE respectively, all around -0.18 and -0.23 K), with a mean absolute error (MAE) of 0.3 and 0.32 K respectively. The low interquantile range (IQR) around 0.37-0.38 K and root mean squared error (RMSE) also around 0.37-0.38 K suggest that the errors are relatively compact and consistent.
On the other hand, the Abilium and Decentlab models exhibit a discrepancy between a positive MBE and a negative MdBE, which indicates a skewed error distribution due to positive outliers. Both Abilium and Decentlab show an error of much larger magnitude (MAE of 0.43 and 0.53 K respectively) and spread (as indicated by the noticeably larger RMSE and IQR values) than the Barani and Koalasense models.
The Onset model shows the smallest errors of all five models (MAE and RMSE of 0.26 K and 0.35 K respectively), with a small and consistent positive bias (MBE and MdBE of 0.08 and 0.04 K respectively), suggesting a moderate yet stable warm offset rather than the radiation-driven skewness observed in the Abilium and Decentlab devices.

\begin{table}[!h]
  \setlength{\tabcolsep}{12pt}
  \begin{adjustwidth}{-.28\textwidth}{0cm}
    \centering
    \begin{threeparttable}
      \footnotesize
      \raggedright
      \begin{tabular}{lrrrrrr}
        \toprule
        \textbf{LCD model} & \textbf{N. samples} & \textbf{MBE [K]} & \textbf{MdBE [K]} &
    \textbf{MAE [K]} & \textbf{RMSE [K]} & \textbf{IQR [K]} \\
        \midrule
        Abilium & 1224 & 0.0659 & -0.0202 & 0.4322 & 0.5490 & 0.7584 \\
        Barani & 1224 & -0.1803 & -0.2333 & 0.3047 & 0.3717 & 0.3667 \\
        Decentlab & 802 & 0.1746 & -0.0769 & 0.5271 & 0.6742 & 1.0258 \\
        Koalasense & 1224 & -0.1878 & -0.2167 & 0.3203 & 0.3870 & 0.3833 \\
        Onset & 1224 & 0.0777 & 0.0418 & 0.2554 & 0.3516 & 0.4106 \\
        \bottomrule
      \end{tabular}
      \caption{
        \label{tab:agreement-metrics} Agreement metrics between each LCD model and the reference AWS.
      }
    \end{threeparttable}
  \end{adjustwidth}
\end{table}

\begin{figure}[!h]
  \centering
  \begin{adjustwidth}{-.4\textwidth}{0cm}
    \includegraphics[width=1.4\textwidth]{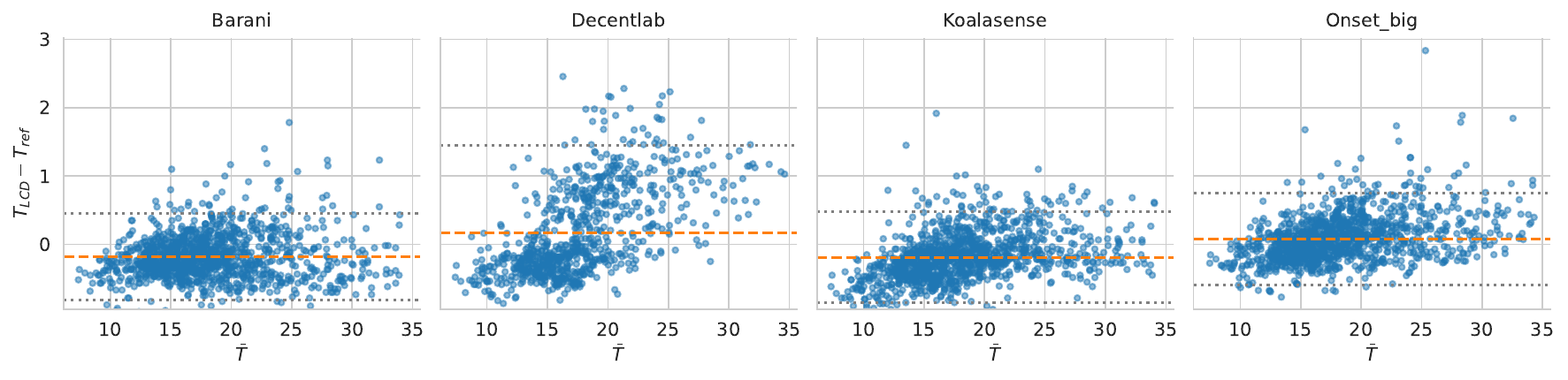}
    \begin{minipage}{\linewidth}
      \caption{\label{fig:bland-altman-plot} Bland-Altman plots comparing each LCD model and the collocated reference AWS during the intercomparison field study. The orange dashed line shows the mean difference ($T_{LCD} - T_{ref}$), whereas the dotted gray lines show the 1.96 standard deviations (SD) limits of agreement.}
    \end{minipage}
  \end{adjustwidth}
\end{figure}

The Bland-Altman plot of \autoref{fig:bland-altman-plot} shows the differences between the measurements of each LCD model and the reference AWS, i.e., $T_{LCD} - T_{AWS}$ against their mean, i.e., $\bar{T} = (T_{LCD} + T_{AWS}) / 2$.
In line with \autoref{tab:agreement-metrics}, The Abilum and Decentlab models show the largest spread in measurement differences, which are positively correlated with the mean, highlighting higher error magnitude at higher temperatures. Such a pattern is not observed for the Barani and Koalansense stations, which show an overall smaller spread in $T_{LCD} - T_{AWS}$ and a more homogenous distribution of errors across the temperature range. Nonetheless, all sensors show an overall asymmetry in the $T_{LCD} - T_{AWS}$ range, with the differences being of higher magnitude for $T_{LCD} > T_{AWS}$. This suggests a skewed error distribution with a higher frequency of positive outliers, especially the Abilium and Decentlab models, which is consistent with their positive MBE and negative MdBE in \autoref{tab:agreement-metrics}.

\subsubsection*{Radiative bias correction performance}

The GAM-based radiative bias correction procedure performs differently for each LCD model (\autoref{tab:radiative-bias-correction}, see the sections ``\nameref{sec:methods-radiative-bias-correction}'' and \nameref{sec:code-radiative-bias} for more details on the statistical methods used).
The Decentlab device shows the largest error reductions in both the MAE (0.30 K, 56.20\%) and RMSE (0.35 K, 52.12\%), followed by Abilium with MAE and RMSE reductions of 39.09\% and 36.49\% respectively. This is consistent with the high proportion ob bias variance explained by the incoming short-wave radiation (R$^2$ of 0.76 and 0.59 respectively).
The Koalasense model achieves intermediate improvements (35.77\% MAE and 30.13\% RMSE reduction, R$^2 =$ 0.36), while the Barani and Onset models show the smallest gains (23.66\% and 12.75\% MAE reduction respectively), suggesting that their differences with the reference AWS are less driven by incoming radiation (R$^2$ of 0.10 and 0.21).

\begin{table}
  \setlength{\tabcolsep}{12pt}
  \begin{adjustwidth}{-.27\textwidth}{0cm}
    \centering
    \begin{threeparttable}
  \footnotesize
  \raggedright
  \begin{tabular}{l r r r r r}
    \toprule
    \textbf{LCD model} & \textbf{R$^2$} & \textbf{MAE [K]} &
  \textbf{RMSE [K]} & \textbf{MAE reduction [K]} & \textbf{RMSE reduction [K]} \\
    \midrule
    Abilium & 0.5907 & 0.2633 & 0.3486 & 0.1690 (39.09\%) & 0.2003 (36.49\%) \\
    Barani & 0.0975 & 0.2326 & 0.3088 & 0.0721 (23.66\%) & 0.0629 (16.93\%) \\
    Decentlab & 0.7543 & 0.2309 & 0.3228 & 0.2962 (56.20\%) & 0.3514 (52.12\%) \\
    Koalasense & 0.3614 & 0.2058 & 0.2704 & 0.1146 (35.77\%) & 0.1166 (30.13\%) \\
    Onset & 0.2136 & 0.2229 & 0.3041 & 0.0326 (12.75\%) & 0.0475 (13.51\%) \\
    \bottomrule
  \end{tabular}    
  \caption{
    \label{tab:radiative-bias-correction}
    Performance metrics of the GAM-based radiative bias correction procedure for each LCD model.
    The MAE and RMSE reductions show the absolute and relative improvements of the correction procedure compared to the raw LCD measurements, which are computed as the difference between the agreement metrics of the raw LCD versus AWS readings (\autoref{tab:agreement-metrics}) and the corrected LCD versus AWS readings.
  }
  \end{threeparttable}
  \end{adjustwidth}
\end{table}



\subsection*{Impacts on heat indices in Swiss cities}
In this section we examine two heat-related indices --- tropical nights and heat warnings --- by comparing the observations from the reference MeteoSwiss AWS and the existing urban climate networks in the cities of Bern, Lausanne, Neuchatel and Zurich (\autoref{sec:study-cases}).
The LCD networks of Bern, Lausanne and Neuchatel use Abilium, Koalasense and Onset sensors respectively, while Zurich comprises two distinct networks: the network of the Office for Waste, Water, Energy and Air of the Canton of Zurich (AWEL), which uses Decentlab sensors, and the network of the Urban Environment and Health Protection Department of the City of Zurich (UGZ), which uses Barani sensors.
For each city, we compare the diurnal temperature cycle, the number of tropical nights as well as the number and magnitude of heat warnings when computing them out of the time series of AWS, raw and corrected LCD measurements, i.e., LCD$_{raw}$ and LCD$_{cor}$ respectively (\nameref{sec:heat-indices}).

\subsubsection*{Diurnal temperature cycle on hot days}

\begin{figure}[!h]
  \centering
  \begin{adjustwidth}{-.4\textwidth}{0cm}  
    \includegraphics[width=1.4\textwidth]{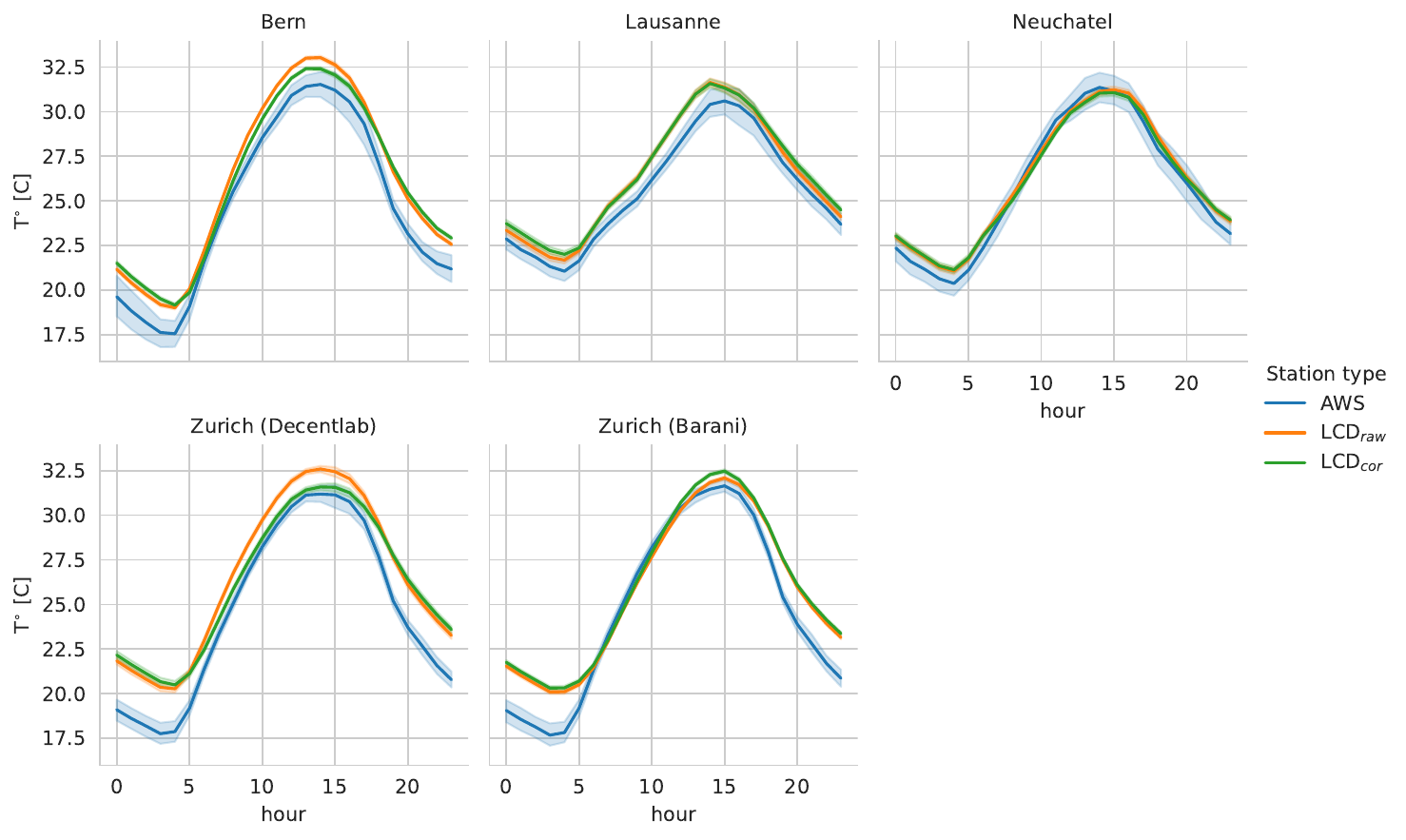}
    \begin{minipage}{\linewidth}    
      \caption{\label{fig:t-diurnal-cycle} Diurnal cycle of temperatures for AWS, LCD$_{raw}$ and LCD$_{cor}$ during hot days with T$_{mean} \geq$ 25 $\degree$C. The lines and shaded areas respectively show the mean and 95\% confidence intervals.}
    \end{minipage}
  \end{adjustwidth}
\end{figure}

In days with T$_{mean} \geq$ 25 $\degree$C, measurements from AWS are consistently lower than their LCD$_{raw}$ counterparts throughout the diurnal temperature cycle (\autoref{fig:t-diurnal-cycle}), even after correcting for radiative errors LCD$_{cor}$.
This pattern is observed for all cities and LCD models, highlighting the underestimation of urban temperatures by AWS located outside of the urban settings, which is consistent with the expected urban heat island effect.
Nevertheless, several idiosyncrasies can be noted, which are likely driven by both the different geographic and physical characteristics of each city as well as by the LCD models used.
In Bern, the difference between the LCD$_{raw}$ and LCD$_{cor}$ is positively correlated with the temperature and radiation cycle, i.e., the difference is negative at nighttime but becomes positive at daytime. The maximum differences are observed at 2 p.m., with an absolute value that is much larger than its nighttime negative difference counterpart.
After applying radiative bias correction, LCD$_{cor}$ suggests that although the maximum diurnal temperatures are likely overestimated by LCD$_{raw}$ in Bern, temperatures in the urban sites are still higher than those measured by the AWS, which are located in less urbanized settings.
A very similar pattern can be observed within the Decentlab network in Zurich, where the differences between LCD$_{raw}$ and LCD$_{cor}$ are further exacerbated during the hottest hours of the day.
On the other hand, the difference between LCD$_{raw}$ and LCD$_{cor}$ within the Barani network in Zurich is very small and consistently negative, i.e., the raw readings LCD$_{raw}$ are overall lower than LCD$_{cor}$.  
A notable detail is that the diural cycle of LCD$_{cor}$ is very similar for both the Decentlab and Barani networks.
In Lausanne, the differences between LCD$_{raw}$ and LCD$_{cor}$ resemble those observed within the Barani network in Zurich, with practically no difference in the maximum temperatures and a small negative difference during nighttime. 
When compared with Bern and Zurich, the differences between LCD$_{cor}$ and AWS observed in Lausanne are also much lower at night, suggesting a smaller nighttime UHI magnitude.
In Neuchatel, the Onset network shows the smallest differences between LCD$_{raw}$ and LCD$_{cor}$ among all cities, with practically no divergence at peak temperatures and only a marginal negative difference at nighttime. The nocturnal offset between LCD and AWS is comparable in magnitude to that observed in Lausanne.
Overall, the differences between LCD and AWS are higher at night, with very little differences between LCD$_{raw}$ and LCD$_{cor}$ due to the absence of incoming short-wave radiation.


\subsubsection*{Tropical nights}

\begin{figure}[!h]
  \centering
    \includegraphics[width=\textwidth]{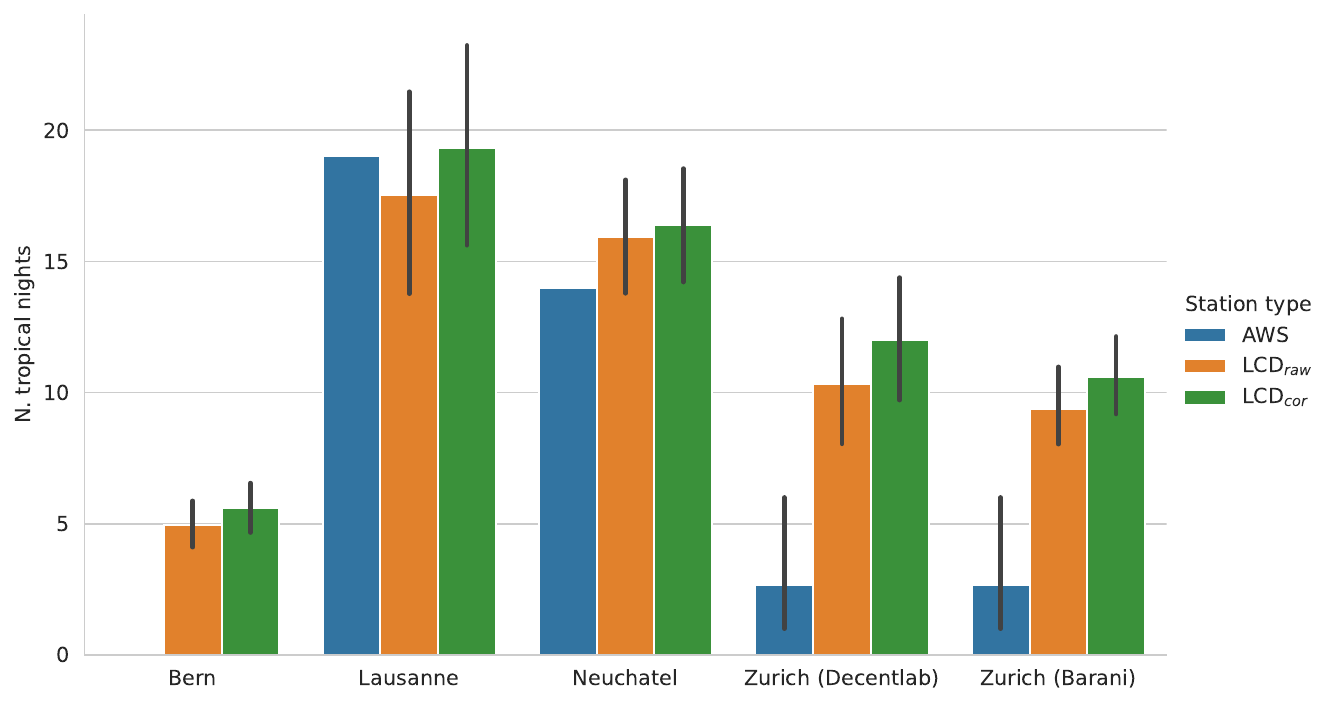}
    \begin{minipage}{\linewidth}    
      \caption{\label{fig:tn-barplot} Number of tropical nights in each city based on AWS, LCD$_{raw}$ and LCD$_{cor}$ data. The bar heights represent the mean number of tropical nights across all stations of each type, whereas the error bars represent the 95\% confidence intervals.}
    \end{minipage}
\end{figure}

\begin{figure}[!p]
  \vspace{-1.5em}
  \centering
  \begin{adjustwidth}{-.4\textwidth}{0cm}
    \includegraphics[width=1.4\textwidth]{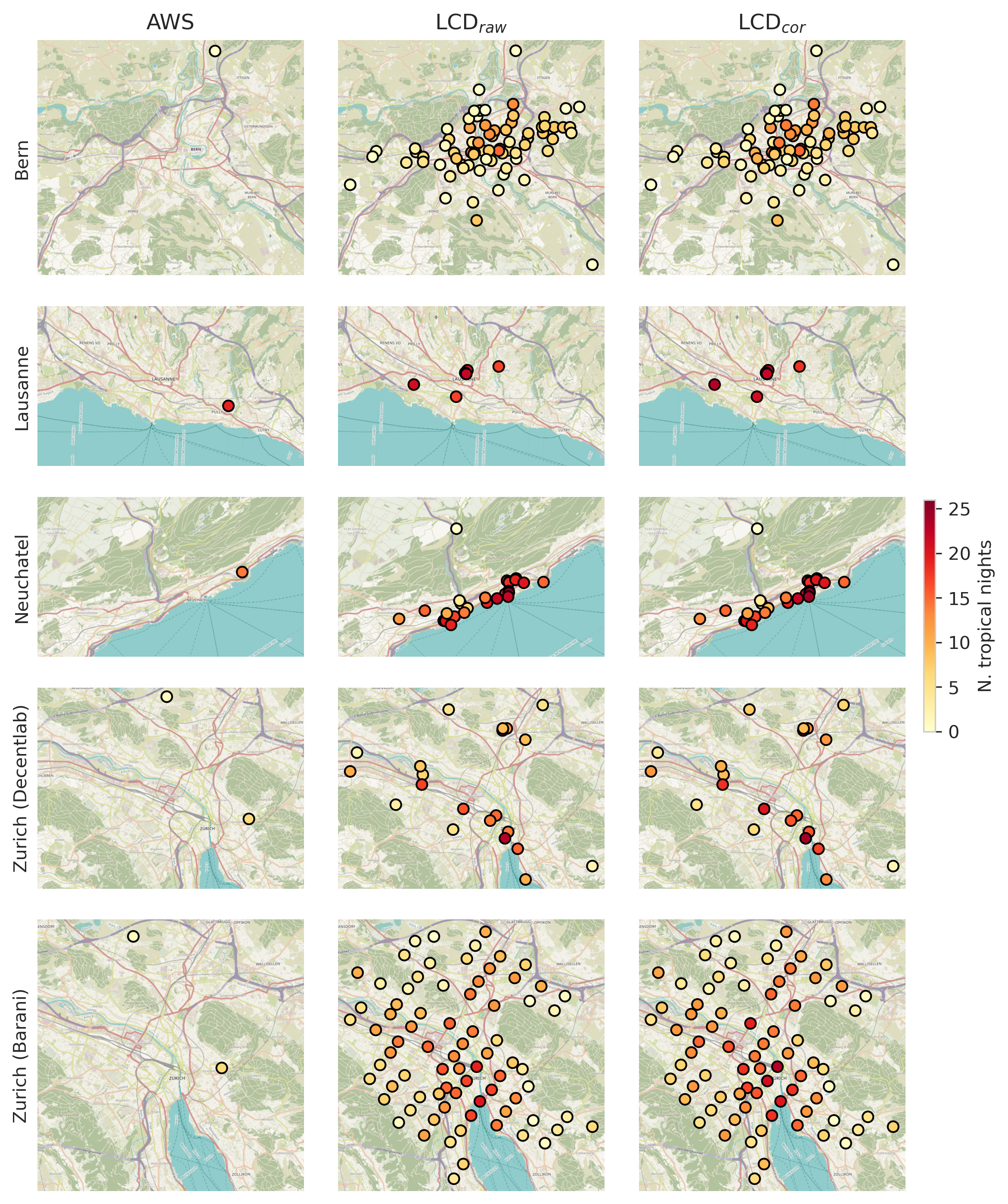}
    \begin{minipage}{\linewidth}
      \caption{
        \label{fig:tn-station-maps}
        Number of tropical nights at each station location, based on AWS, LCD$_{raw}$ and LCD$_{cor}$ data.
      }
    \end{minipage}
  \end{adjustwidth}
\end{figure}

The number of tropical nights (i.e., nights with T$_{min} \geq$ 20 $\degree$C) is notably higher when computed with LCD data than with AWS, which suggests that the latter may be underestimating the number of tropical nights in the urbanized zones (\autoref{fig:tn-barplot}, \autoref{fig:tn-station-maps}).

In Bern, no tropical nights occur within the study period according to the AWS data\footnote{As of the time of writing, no tropical night has ever been recorded at Bern/Zollikofen (BER).}, whereas averaging across all LCD stations yields 4.96 and 5.56 tropical nights for LCD$_{raw}$ and LCD$_{cor}$ respectively.
Given that Zurich is the only study area with more than one AWS, the corresponding number of tropical nights ranges from 1 to 6, with an average of 2.67. When considering the Decentlab network, the average number of tropical nights at each site is 10.33 for LCD$_{raw}$ and 12 for LCD$_{cor}$, which is slightly higher than in the Barani network (9.38 and 10.58 respectively).
In Neuchatel, 14 tropical nights are reported according to the AWS, which is slightly below the mean across Onset stations (15.93 for LCD$_{raw}$ and 16.36 for LCD$_{cor}$).
In Lausanne, the AWS reports 19 tropical nights, which is higher than the average out LCD station readings LCD$_{raw}$ (17.54) but slightly lower than the mean accross corrected values LCD$_{cor}$ (19.31).
As observed for the diurnal cycle \autoref{fig:t-diurnal-cycle}, the smaller differences between AWS and LCD partly attributable to the large elevation gradient in Lausanne. Nonetheless, at first glance, the values reported at each station location does not clearly support this hypothesis (\autoref{fig:tn-station-maps}), which may instead point to more complex local topographic effects.
On the other hand, the spatial pattern of the number of tropical nights in Zurich, Bern (and to a lesser extent Neuchatel) seems to be more consistent with the expected UHI pattern (\autoref{fig:tn-station-maps}).

\subsubsection*{MeteoSwiss heat warning levels}

\begin{figure}[!h]
  \centering
  \begin{adjustwidth}{-.4\textwidth}{0cm}  
    \includegraphics[width=1.4\textwidth]{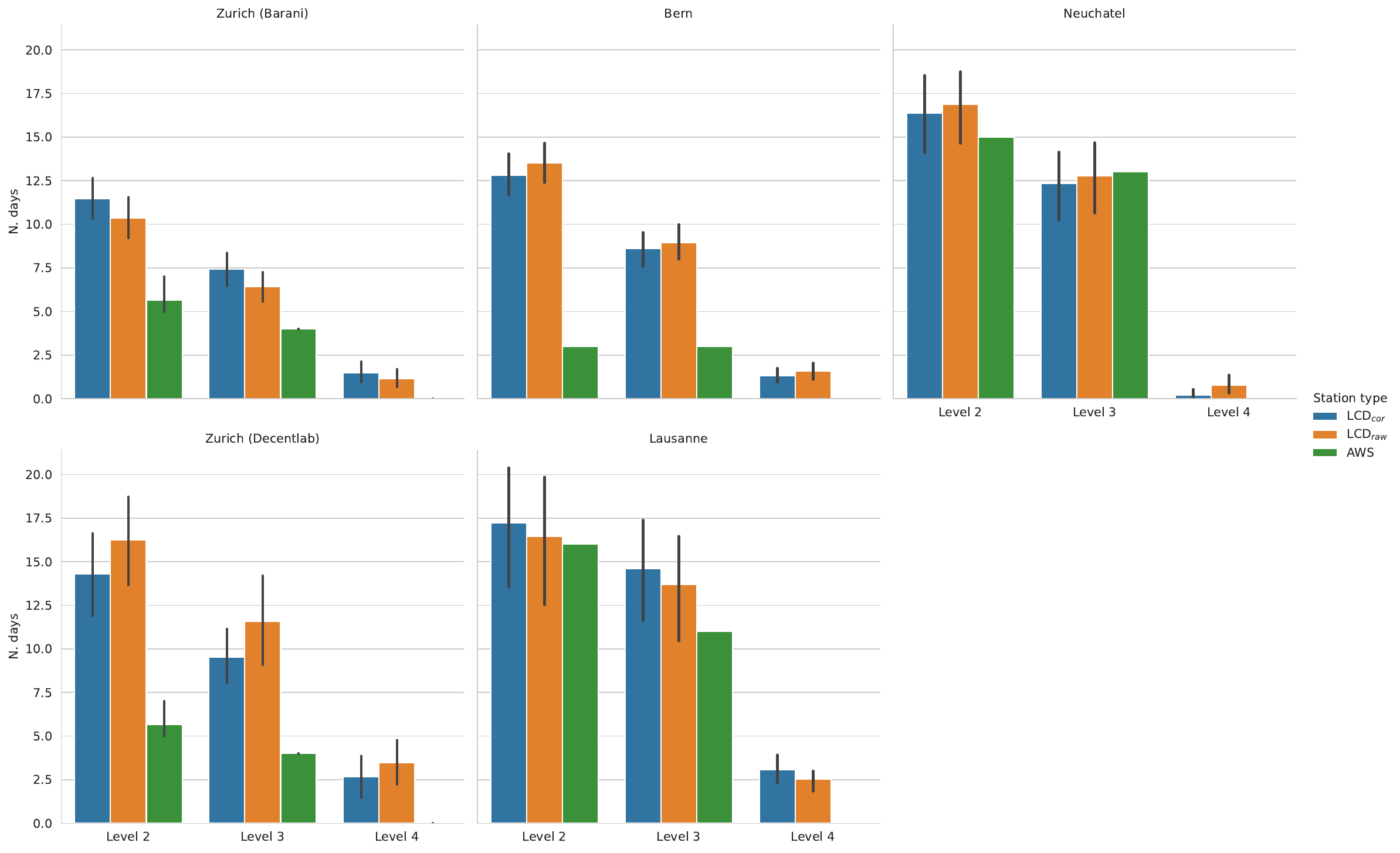}
    \begin{minipage}{\linewidth}    
      \caption{\label{fig:heat-warnings-barplot} Number of days under each heat warning based on AWS, LCD$_{raw}$ and LCD$_{cor}$ data.}
    \end{minipage}
  \end{adjustwidth}
\end{figure}

As shown in \autoref{fig:heat-warnings-barplot}, LCD stations consistently report more hot days than the reference AWS for all heat-warning levels, with the largest gaps in Bern and Zurich (especially for the Decentlab network).
In Bern, the AWS reports 3 days under level 2 and 3 warnings and zero for level 4 whereas the mean day counts across LCD stations are much higher and very similar before and after correction: 13.52 vs 12.60 (level 2), 8.94 vs 8.51 (level 3), and 1.58 vs 1.30 (level 4) for LCD$_{raw}$ and LCD$_{cor}$ respectively. This suggests that estimates based on raw and corrected LCD readings are close to each other while remaining far above the AWS values.
In Zurich, both LCD networks report higher day counts for all levels, with Decentlab consistently higher than Barani both before and after correction: for level 2 warnings, 16.24/14.29 (Decentlab LCD$_{raw}$/LCD$_{cor}$) and 10.36/11.55 (Barani) versus 5.67 days according to the AWS; for level 3, 11.57/9.52 (Decentlab) and 6.44/7.56 (Barani) versus 4 (AWS); and for level 4, Decentlab reports 3.48/2.67 and Barani 1.16/1.50 versus 0 days according to AWS.
Lausanne shows the smallest differences between LCD and AWS, with 16.46 and 17.15 days of level 2 warnings for LCD$_{raw}$ and LCD$_{cor}$ respectively versus 16 according to AWS, then 13.69 and 14.31 versus 11 for level 3 and 2.54 and 3.00 versus 0.
In Neuchatel, the differences between LCD and AWS are the smallest of all cities: for level 2 warnings, LCD$_{raw}$ and LCD$_{cor}$ report 16.89 and 16.43 days respectively versus 15 according to the AWS; for level 3, LCD reported days are slightly below AWS (12.79 and 12.36 versus 13); and level 4 warnings are practically absent for both LCD and AWS 0.79 and 0.21 versus 0).
The correction procedure impacts heat warnings differently: it reduces counts in Bern, Neuchatel and for the Zurich Decentlab network but increases them in Lausanne and the Zurich Barani network.

\section*{Discussion}

This study shows that the estimated UHI magnitude in Swiss cities depends strongly on the observation network used. Across our study cases in Bern, Lausanne, Neuchatel and Zurich, AWS observations are generally cooler than its LCD counterparts during hot conditions, a difference which propagates to impactful heat indices such as the number of tropical nights and heat warnings issued.

A key question that arises is to what extent these differences can be mostly attributed to radiative errors of LCDs or they instead reflect actual hotter conditions experienced in urban settings. Despite notable differences between LCD models, the evidence presented in this article strongly supports the latter.
First, the agreement metrics for the intercomparison field study show that the Barani and Koalasense sensors present compact and stable differences with the professional AWS operated by MeteoSwiss. Furthermore, the differences of between these LCD models and the AWS are actually negative as reported by both the MBE and MdBE, which suggests that the LCD readings are overall cooler than the reference AWS.
Despite this, the Barani and Koalasense readings within their respective LCD networks in Zurich and Lausanne report higher temperatures than their nearest AWS throughout the diurnal cycle.
The Onset device also shows compact and stable errors, albeit with a small positive bias; yet the Onset network in Neuchatel still reports consistently more tropical nights than the reference AWS.

Secondly, while the Abilium and Decentlab LCD models do exhibit a pattern of radiation-driven positive biases, the GAM-based correction procedure significantly reduces the errors of these models, resulting in corrected readings that are overall closer to the reference AWS. However, even after correction, the number of heat warnings based on LCD$_{cor}$ remain notably higher than those based on AWS, which suggests that the differences between the two networks cannot be solely attributed to radiative errors.
Additionally, radiative errors should have little impact at nighttime yet the number of tropical nights based on LCD data is still significantly higher than reported by AWS.
Similarly, despite notable differences in the agreement between the reference AWS and the Decentlab and Barani LCD models, the diurnal cycle and number of tropical nights based on LCD$_{cor}$ data from the respective LCD networks in Zurich is remarkably similar.
This suggests that while the LCD$_{raw}$ data from Decentlab may overestimate hot conditions, temperatures at the corresponding urban LCD sites are still significantly higher than reported by the nearest AWS, which are located in less urbanized settings.

Finally, the spatial patterns of the number of tropical nights in Bern and Zurich seem to be consistent with the expected UHI pattern, with higher values in the more urbanized zones and lower values in the less urbanized zones.
In contrast, both Lausanne and Neuchatel show much smaller differences between LCD and AWS across all heat indices.
Notably, the Onset network in Neuchatel (i.e., the best-performing sensor in the intercomparison field study) still reports consistently more tropical nights than the reference AWS, which suggests that the small but persistent LCD-AWS gap reflects a real urban warming signal rather than solely sensor biases.
The reversal of level 3 heat warning counts in Neuchatel (where LCD falls below AWS) and the near-identical tropical night counts in Lausanne may instead point to more complex local topographic effects in both cities, such as cold-air drainage, valley-channelled winds, slope and aspect effects on radiation or persistent nocturnal inversions.

Overall, the evidence presented in this article suggests that, despite notable radiative biases in some LCD models, professional AWS located outside urban settings underestimate the heat risks experienced within urbanized zones.
At the same time, our analysis identifies sensor-specific limitations that can directly inform the selection and deployment of new LCD models to improve urban climate networks, and provides a practical radiative bias correction approach for existing networks using sensors prone to this issue.
Most importantly, we show that lower-performing LCDs can provide more representative estimates of urban heat exposure than AWS stations located outside the urban core.

The nocturnal UHI pattern observed across all four cities is consistent with the energetic basis of the urban heat island effect \citep{oke1982energetic} and aligns with previous analysis of the UHI effect in Swiss cities using paired rural and urban AWS \citep{burgstall2019representing}.
The spatial distribution of intra-urban temperatures in Bern is further consistent with land use regression models of nighttime temperatures derived from the same Bern LCD network \citep{burger2021modelling,burger2022modeling}, in which built-up land cover and cold air drainage emerged as the dominant warming and cooling drivers.
However, those studies relied on nighttime only data because of daytime radiative biases in the Bern LCDs previously identified by \citet{gubler2021evaluation} --- which the GAM-based approach presented here addresses.
Similarly, the intra-urban temperature pattern in Zurich is consistent with machine-learning-based urban temperature mapping using the same AWEL Decentlab network alongside Netatmo citizen weather stations \citep{zumwald2021mapping}, in which building density and vegetation cover emerged as the dominant spatial drivers of intra-urban temperature variability.
Finally, the smaller LCD-AWS differences in Lausanne and Neuchatel are consistent with more complex topographic settings; in Lausanne, previous work found no significant elevation effect on nocturnal AWS temperatures \citep{bosch2021spatially}, which may reflect the limited spatial coverage of the existing AWS network rather than an absence of topographic influence.

Several limitations of this study are worth noting. First, the intercomparison field study was conducted at a single rural, open-sky site, nevertheless urban installations may be subject to different radiation environments due to building shading and wall-reflected radiation. These differences could affect both the magnitude of radiative biases and the transferability of the GAM-based correction to deployed urban sensors.
Second, the analysis covers a single summer season, so long-term sensor stability, degradation effects and inter-annual variability of the radiative bias behaviour remain unknown.
Finally, the AWS stations used as reference are located outside urban areas, meaning that the LCD-AWS gap conflates actual urban warming with potential differences in station siting. However, the convergence of Decentlab and Barani corrected readings in Zurich, as well as the persistent LCD-AWS gap for the well-performing Onset sensors in Neuchatel, suggest that siting alone cannot explain the observed differences.

\section*{Materials and methods}

\subsection*{Intercomparison field study}
\label{sec:interc-field-study}

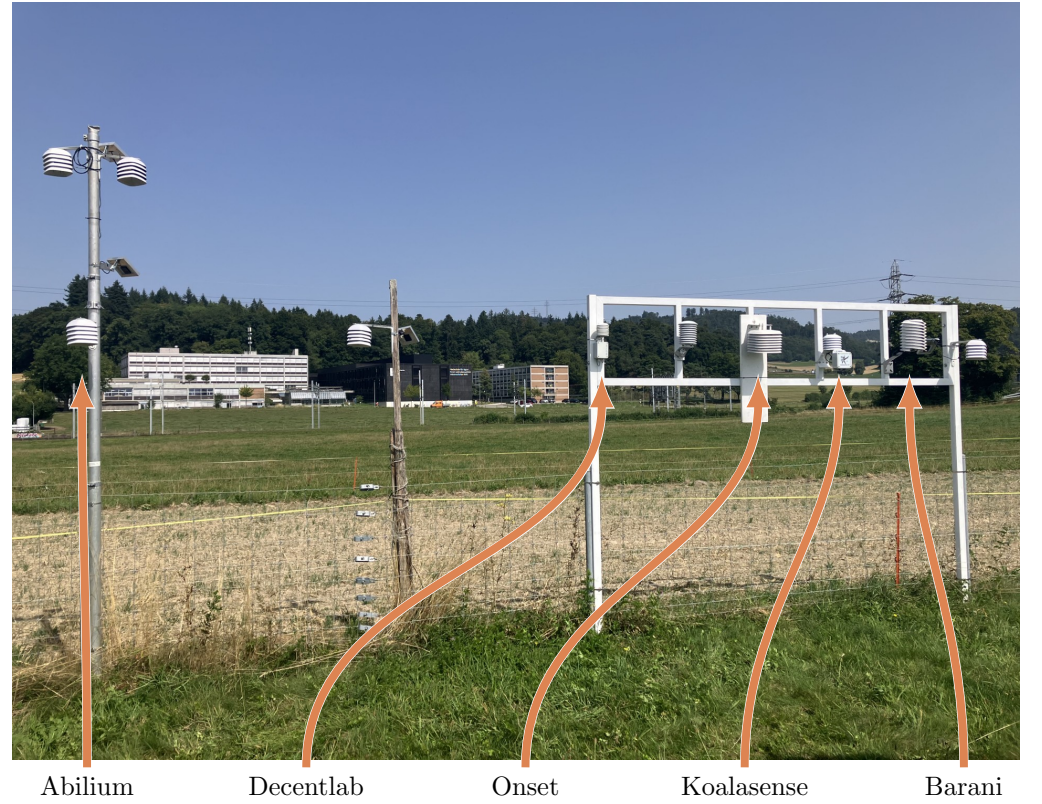
\begin{figure}[!h]
  \centering
  \begin{tikzpicture}
    \node [anchor=north] (abilium) at (1,1.4) {Abilium};
    \node [anchor=north] (decentlab) at (3.9,1.4) {Decentlab};
    \node [anchor=north] (onset) at (6.8,1.4) {Onset};
    \node [anchor=north] (koalasense) at (9.7,1.4) {Koalasense};
    \node [anchor=north] (barani) at (12.6,1.4) {Barani};    
    \begin{scope}[yshift=1.5cm]
      \node[anchor=south west,inner sep=0] (image) at (0,0) {\includegraphics[width=\textwidth]{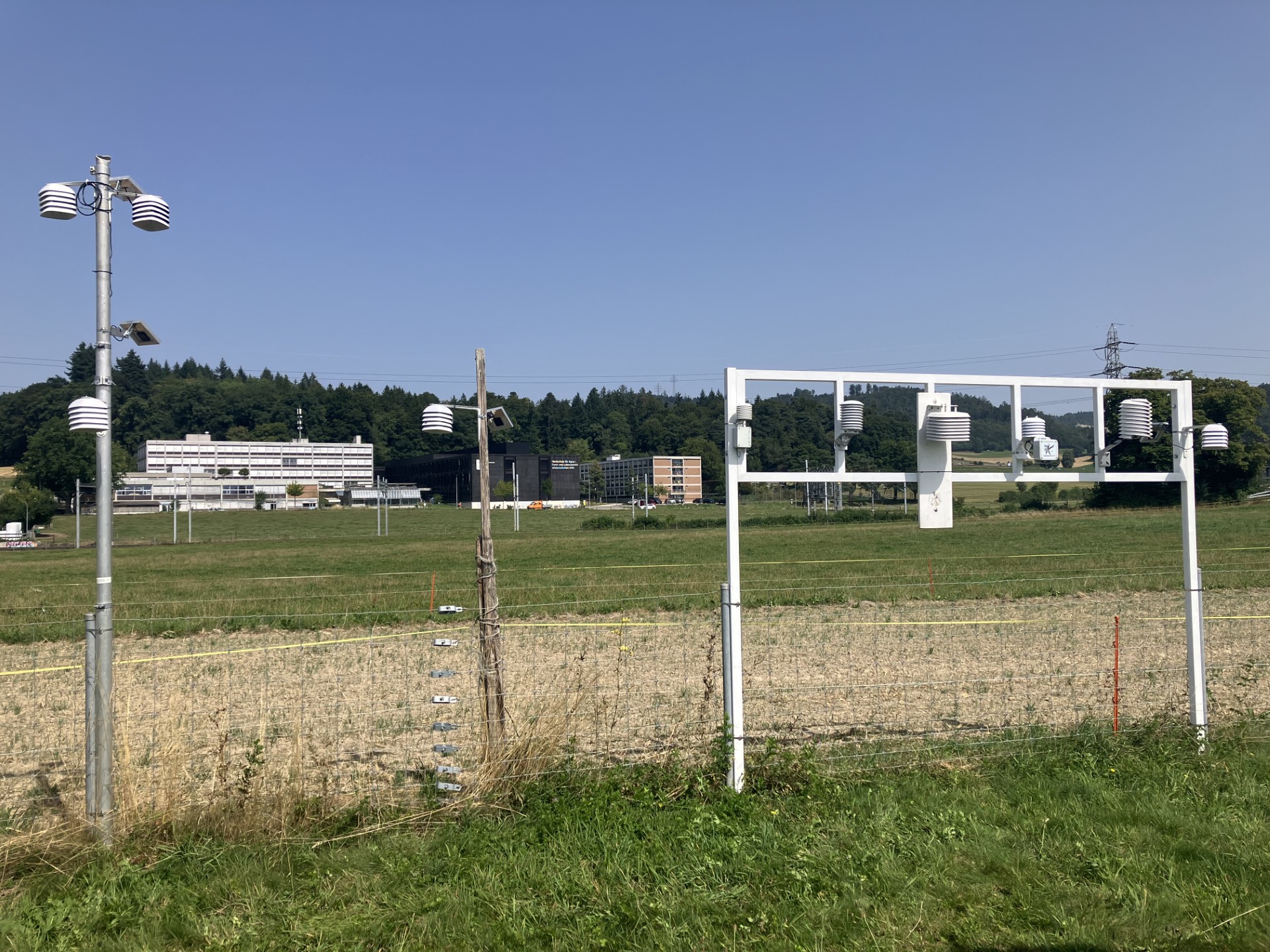}};
      \begin{scope}[x={(image.south east)},y={(image.north west)}]
        \newcommand{\arrowWithBorder}[5]{%
          \draw[
          #5,
          line width=#4+.5pt,
          white,
          ] (#1) to[out=90,in=-90] #2;
          \draw[
          #5,
          line width=#4,
          shorten >=1pt,          
          #3
          ] (#1) to[out=90,in=-90] #2;
        }
        \arrowWithBorder{abilium}{(0.07,0.51)}{sns-orange}{3pt}{-latex}
        \arrowWithBorder{decentlab}{(0.585,0.51)}{sns-orange}{3pt}{-latex}
        \arrowWithBorder{onset}{(0.74,0.51)}{sns-orange}{3pt}{-latex}
        \arrowWithBorder{koalasense}{(0.82,0.51)}{sns-orange}{3pt}{-latex}
        \arrowWithBorder{barani}{(0.89,0.51)}{sns-orange}{3pt}{-latex}        
      \end{scope}
    \end{scope}
  \end{tikzpicture}%
  \vspace{.5em}
  \caption{\label{fig:intercomparison-field-setup-zollikofen} Set up for the intercomparison field study.}
\end{figure}

The intercomparison field study was performed during 7 weeks in summer 2025 (from July 27 to September 15) using seven LCD models collocated alongside a reference professional AWS at Zollikofen (\autoref{fig:intercomparison-field-setup-zollikofen}), which is part of the automated observation network of the Swiss Federal Office of Meteorology and Climatology (MeteoSwiss) and is situated about 5 km north of the Bern city center in a rural environment.

The main characteristics of the LCDs and the MeteoSwiss AWS are listed in \autoref{tab:sensors}.

\begin{table}
  \setlength{\tabcolsep}{12pt}
  \begin{adjustwidth}{-.4\textwidth}{0cm}
    \centering
    \begin{threeparttable}
      \footnotesize
      \raggedright
      \begin{tabular}{p{.10\textwidth} p{.20\textwidth} p{.18\textwidth} p{.14\textwidth} p{.28\textwidth} p{.10\textwidth}}
        \toprule
        \textbf{Name} & \textbf{Sensor Model} & \textbf{Manufacturer} & \textbf{Accuracy (K)} & \textbf{Other} & \textbf{References} \\
        \midrule
        Albilium & Sensirion SHT31A-DIS-B2.5kS & Sensirion AG & $\pm$ 0.3 (-40 to 90 $\degree$C) & Five white melamine formaldehyde resin plates & \cite{hari2022development} \\
        Decentlab & Sensirion SHT35 & Sensirion AG & $\pm$ 0.2 (-40 to 90 $\degree$C) & Passive radiation shield & \cite{decentlab2018data} \\
        Barani & MeteoHelix\textsuperscript{\textregistered} IoT Pro & BARANI DESIGN Technologies s.r.o. & $\pm$ 0.1 (0 to 65 $\degree$C) & Naturally ventilated double-helix radiation shield (MeteoShield\textsuperscript{\textregistered} Pro) & \cite{barani2024data} \\
        MeteoSwiss & Thygan VTP6 & Meteolabor AG & $\pm$ 0.15 (-20 to 50 $\degree$C) & Actively ventilated & \cite{meteolabor2019data} \\
        \bottomrule
      \end{tabular}
      \caption{Technical specifications of the LCD models and reference AWS used in the intercomparison field study. Technical specifications for the Koalasense model were not available from the manufacturer.}
      \label{tab:sensors}
    \end{threeparttable}
  \end{adjustwidth}
\end{table}

\subsubsection*{Agreement metrics}

For each LCD model, the agreement with the reference AWS was quantified from concurrent hourly temperature pairs. Given the temperature difference at time \(t\) and \(n\) the number of valid paired observations for each LCD model, defined as:

\[
  e_t = T^{LCD}_t - T^{ref}_t
\]

we report the following agreement metrics:

\[
  \mathrm{MBE} = \frac{1}{n}\sum_{t=1}^{n} e_t,\quad
  \mathrm{MdBE} = \mathrm{median}(e_t),\quad
  \mathrm{MAE} = \frac{1}{n}\sum_{t=1}^{n} |e_t|,
\]
\[
  \mathrm{RMSE} = \sqrt{\frac{1}{n}\sum_{t=1}^{n} e_t^2},\quad
  \mathrm{IQR} = Q_{0.75}(e_t)-Q_{0.25}(e_t).
\]

Positive biases indicate warmer LCD readings than the reference AWS. The MBE and RMSE summarize average and squared-error behavior respectively, while MdBE and IQR provide robust metrics less sensitive to skewness and outliers.

Agreement was additionally assessed with Bland-Altman plots \cite{altman1983measurement,bland1986statistical}, which have been widely used to quantify differences between air temperature sensors \cite{dons2017wearable,ueberham2018wearable,bailey2020wearable,sugg2022individually}.

\subsubsection*{Radiative bias correction}
\label{sec:methods-radiative-bias-correction}

In line with related works \cite{jenkins2014comparison,bell2015good,buchau2018modelling,cornes2020correcting,gubler2021evaluation,beele2022quality,anet2024improving,ahmed2025comparison}, we model the temperature difference between the LCDs and the AWS ($\Delta T_t$ at time $t$) and the short wave radiation ($Rad_t$) using a generalized additive modelling (GAM) approach (see \nameref{sec:code-radiative-bias}) as in:

\begin{equation}
  \Delta T_t \sim \beta_0 + f ( Rad_t ) + \epsilon_t  
\end{equation}

where $\beta_0$ represents the intercept term, $f$ is a smoothing function applied to the radiation term and $\epsilon_t$ corresponds to the model residuals\footnote{Following \citet{cornes2020correcting}, we additionally tested an auto-regressive term to account for temporal autocorrelation in the residuals. Beyond not resulting in a meaningful improvement of the statistical fit, the auto-regressive term requires observed $\Delta T_{t-k}$ values at prediction time, which are unavailable when applying the correction to deployed LCD stations where no concurrent reference measurement exists (see \nameref{sec:code-radiative-bias}).}

Then, given $Rad_t$, the corrected temperature can be estimated as in:

\begin{equation}
  \hat{T}_t^{LCD} = T_t^{LCD} - \Delta T_t
\end{equation}

For each LCD, $Rad_t$ is computed as a rolling sum of the short-wave radiation from the nearest AWS, with the time window (1 to 5 hours) selected per station by maximizing the Pearson correlation with $\Delta T_t$.

The smoothing function $f$ is fitted as an Explainable Boosting Machine (EBM) \cite{lou2013accurate}, a class of GAM in which shape functions are learned via cyclic gradient boosting rather than penalized regression splines, as provided by the InterpretML library \citep{nori2019interpretml}.

\subsection*{Study cases and measurement networks}
\label{sec:study-cases}

This study examines the urban climate networks of the Swiss cities of Bern, Lausanne, Neuchatel and Zurich, whose main characteristics are summarized in \autoref{tab:study-cases} (see also \autoref{tab:sensors} for details on the LCD models).
All reference stations are part of the MeteoSwiss automated observation network (SwissMetNet).
In Bern, the LCD network consists of Abilium sensors deployed since summer 2018 \citep{gubler2021evaluation}.
In Lausanne and Neuchatel, the respective communal administrations operate networks of Koalasense and Onset sensors.
In Zurich, two separate networks are considered: the AWEL network of Decentlab sensors operated by the cantonal environmental agency \citep{awel2026lufttemperatur} and the UGZ network of Barani sensors operated by the city's environmental protection office \citep{ugz2026stadtklima}.
For each network, we selected the nearest SwissMetNet station as reference: Bern/Zollikofen (BER) for Bern --- the same station used in the intercomparison field study ---, Pully (PUY) for Lausanne, Neuch\^atel (NEU) for Neuchatel, and Z\"urich/Affoltern (REH) and Z\"urich/Fluntern (SMA) for Zurich.
In addition to temperature, global radiation (incoming shortwave radiation, routinely measured at SwissMetNet stations) was used as the radiation input $Rad_t$ for the bias correction procedure.

\begin{table}
  \setlength{\tabcolsep}{12pt}
  \begin{adjustwidth}{-.2\textwidth}{0cm}
    \centering
    \begin{threeparttable}
      \footnotesize
      \raggedright
      \begin{tabular}{p{.15\textwidth} p{.15\textwidth} p{.2\textwidth} p{.2\textwidth} p{.10\textwidth}}
        \toprule
        \textbf{City} & \textbf{LCD model} & \textbf{N. valid LCDs\footnote{A station is considered valid if reporting at least 80\% of valid measurements during June, July and August of 2025}} & \textbf{Reference station} & \textbf{References} \\
        \midrule
        Bern & Abilium & 77 & BER & \cite{gubler2021evaluation,burger2026urs} \\
        Lausanne & Koalasense & 7 & PUY & \cite{burger2026urs} \\
        Neuchatel & Onset & 28 & NEU & \cite{burger2026urs} \\
        Zurich (AWEL) & Decentlab & 25 & REH, SMA & \cite{awel2026lufttemperatur} \\
        Zurich (UGZ) & Barani & 83 & REH, SMA & \cite{ugz2026stadtklima,burger2026urs} \\
        \bottomrule
      \end{tabular}
      \caption{Urban climate networks and reference AWS for each study case.}
      \label{tab:study-cases}
    \end{threeparttable}
  \end{adjustwidth}
\end{table}

\section*{Acknowledgements}

The authors thank Patrick Kallabis for building the frame for the parallel measurements in Bern-Zollikofen.

\section*{Funding}

This work was funded by the MeteoSwiss GAW-GCOS Project ``Upscaling Low-cost Urban Climate Networks with Stakeholder Needs'' (ULUC).

\section*{Supplementary Material}

\setcounter{figure}{0}
\renewcommand{\thefigure}{S\arabic{figure}}

\subsection*{Code S1}
\label{sec:code-agreement-metrics}
Agreement metrics between each LCD model and the reference AWS, as a Jupyter notebook (IPYNB).
\href{https://github.com/martibosch/swiss-uhi-lcd/blob/main/notebooks/agreement-metrics.ipynb}{github.com/martibosch/swiss-uhi-lcd/blob/main/notebooks/agreement-metrics.ipynb}

\subsection*{Code S2}
\label{sec:code-radiative-bias}
Training of the GAM-based radiative bias correction model for each LCD, as a Jupyter notebook (IPYNB).
\href{https://github.com/martibosch/swiss-uhi-lcd/blob/main/notebooks/train-bias-correction.ipynb}{github.com/martibosch/swiss-uhi-lcd/blob/main/notebooks/train-bias-correction.ipynb}

\subsection*{Code S3}
\label{sec:code-apply-bias}
Application of the trained GAM correction models to the LCD measurements and validation of correction performance, as a Jupyter notebook (IPYNB).
\href{https://github.com/martibosch/swiss-uhi-lcd/blob/main/notebooks/apply-bias-correction.ipynb}{github.com/martibosch/swiss-uhi-lcd/blob/main/notebooks/apply-bias-correction.ipynb}

\subsection*{Code S4}
\label{sec:heat-indices}
Heat indices computed using data from AWS, LCD$_{raw}$ and LCD$_{cor}$, as a Jupyter notebook (IPYNB).
\href{https://github.com/martibosch/swiss-uhi-lcd/blob/main/notebooks/heat-indicds.ipynb}{github.com/martibosch/swiss-uhi-lcd/blob/main/notebooks/heat-indices.ipynb}

\subsection*{Model weights S1}
\label{sec:model-weights}
Trained GAM-based radiative bias correction models for each LCD model, as serialized model files (SKOPS) hosted on the Hugging Face Hub.
\href{https://huggingface.co/martibosch/lcd-bias-correction}{huggingface.co/martibosch/lcd-bias-correction}


\bibliography{references}

\bibliographystyle{unsrtnat}

\end{document}